\documentclass[showpacs,floatfix,preprint,nofootinbib]{revtex4}
\usepackage{graphics}
\usepackage{bm}

\let\D=\Delta

\let\l=\lambda

\begin{document}
\title{Approach for modelling quantum-mechanical  collapse}
\author{A.~Yu.~Ignatiev}
\email{a.ignatiev@ritp.org}
 \affiliation {{\em Theoretical Physics Research Institute,
     Melbourne 3163,}\\
    {\em   Australia}}
\def\be{\begin{equation}}
\def\ee{\end{equation}}
\def\bea{\begin{eqnarray}}
\def\eea{\end{eqnarray}}
\newcommand{\nn}{\nonumber \\}
\begin{abstract}
A long-standing  quantum-mechanical puzzle is whether the collapse of the wave function is a real physical process or simply an epiphenomenon. This puzzle lies at the heart of the measurement problem.
One way to choose between the alternatives is to assume that one or the other is correct and attempt to draw physical, observable consequences which then could be empirically verified or ruled out.

As a working hypothesis, we propose simple models of collapse as a real physical process for direct binary symmetric measurements made on one particle.
This allows one to construct irreversible unstable
Schr\"odinger equations   capable of describing 
continuously the  process of collapse induced by the interaction of the  quantum system with the measuring device.  Due to unknown initial conditions the collapse outcome remains unpredictable  so no contradictions with quantum mechanics arise.

Our theoretical framework  predicts a finite time-scale of the collapse and links with experiment.
Sensitive probes of the 
 collapse dynamics could be done using Bose-Einstein condensates, 
ultracold neutrons or ultrafast optics.
 If confirmed, the  formulation could be relevant to the transition from quantum fluctuations to classical inhomogeneities in early cosmology and to establishing the ultimate limits on the speed of quantum computation and information processing.
 \end{abstract}
\maketitle

\section{Introduction} 
Modern technology has turned the in-depth studies of quantum mechanics into a fast-growing research area.
The physical meaning  and the evolution of quantum state are among the focal points  in these studies \cite{l,vN,pbr}.  

The wave function collapse (also known as the reduction of the wave packet) is
 an irreversible, unpredictable change of quantum state which occurs during the 
 measurement and which cannot be described by the reversible Schr\"odinger 
 equation.  The nature and physical mechanisms of collapse are among the most debatable issues of quantum mechanics, and  our understanding of this mysterious process remains rather limited. 
as the Many-Worlds interpretation of quantum mechanics.

Collapse has been discussed from numerous, often very different perspectives, starting from purely philosophical and ending with purely mathematical. Philosophical discussions  focus on various interpretations of quantum mechanics, and point out that in some of them, collapse is not a physical process but rather a (trivial) mental process of updating the available information. 

This approach is often illustrated with the example of picking  balls from a box with one black and one white ball. Initially, the probability to pick the white ball is 1/2. But after  the first picked ball turns out black, then the probability to pick up the white ball becomes 1. In this case, which is completely classical, even the term `collapse' itself seems unnecessary. Indeed, before the advent of quantum mechanics, no-one was talking about collapses in the theory of probability.

Such a viewpoint emerged when the options for quantum experimenting were limited by the level of technology.  Since then,  completely new kinds of experiments became a reality, and it is now possible to consider  the collapse problem afresh.

Besides, the quantum collapse without observers has become a vital issue in cosmology where a physical mechanism is needed for turning quantum fluctuations into classical which would then evolve into galaxies.

As a consequence, the modern approach  consists in translating philosophical questions into physical. In this vein, known as ``experimental metaphysics'', we do not postulate `absence of collapse' or otherwise, but rather ask how we could find out if there is a collapse or not using ordinary physical methods and principles. 

The problem of the collapse under measurement is intimately related (but not identical) to the problem of quantum-to-classical transition or  the transition from possible to actual.

The theory of environmental decoherence has made impressive progress in this area over recent decades. Optimistically, a view was advocated by some authors that decoherence has solved the measurement problem and the concept of collapse has become superfluous.  

However, this issue is far from being settled (see, e.g., \cite{adler} and references therein). 

The conventional quantum mechanics models the transition from quantum to classical as a black box.

The decoherence theory has shone some light on that box, but do we know what is inside? What happens at the final, most intriguing stage of the measurement process? Only 15 percent of experts believe that the decoherence program has solved the measurement problem \cite{SKZ}.

For this reason, it seems warranted to consider first the case of `pure' collapse, i.e., neglect the environmental decoherence altogether. This approach can be justified in the end by comparing the two characteristic time-scales: for collapse and for decoherence. If the former is much shorter than the latter, then decoherence indeed can be neglected. This case is also more interesting from a practical point of view as in quantum computing decoherence needs to be reduced as much as possible.


\section{Non-linear effective equations for collapse}
The problem of collapse during the measurement is essentially a many body problem (or even an infinitely many degrees of freedom problem).  For this reason,  exact equations governing the collapse process are hard to obtain, and even if obtained, would be difficult to use.
In such cases, a  common method to proceed is  to eliminate  the unwanted degrees of freedom by applying the phenomenological, effective theory approach. The essence of this approach is to `integrate out' the unknown quantities. This means introducing new terms built out of the known quantities subject to certain symmetry requirements and multiplying them by unknown coefficients. This has become a standard practice of quantum  field theory over the recent decades but is less appreciated in the context of quantum mechanics.

When introducing new terms, our strategy is coservative, and  we will only rely  on the most general physical criteria: instability, irreversibility, and symmetry breaking. The first two features have been advocated already by Niels Bohr, while the third one is among the most universal phenomena in contemporary physics. 

 We  start with the familiar setting where collapse is supposed to occur: a quantum system with a wave function plus a measuring apparatus.

 Following the spirit of the effective-theory, phenomenological type of  approach,  
 we set ourselves the following task: Take the Schr\"odinger equation and add new terms which are made entirely out of the wave function and unknown coefficients. The new equations should be able to describe both ordinary evolution and the measurement process (i.e., the collapse of the wave function).

This may sound as an extraordinary proposal which is bound to fail. Indeed,  it seems that nothing could be so different as the continuous, deterministic evolution and the sudden, unpredictable collapse. Further, it seems unavoidable that some random, stochastic sources, or an additional physical phenomena should necessarily be introduced to account for the collapse as was done, for instance, in Refs.\cite{grw,p89,w11,pen,dio}. 

 gravity-induced collapse ideas of Penrose \cite{pen} and Di\'osi \cite{dio}.

Nevertheless,  we require  from the outset that no new stochastic sources or gravitation should be invoked in explaining the collapse \footnote{To avoid confusion, we emphasize that the Continuous Spontaneous Localization (CSL) theory of  Ghirardi- 
Rimini-Weber and Pearle \cite{grw,p89,w11} and the gravity-induced collapse models, for instance, those  of Penrose \cite{pen} and Di\'osi \cite{dio} are in principle different from our proposal because they address an essentially different problem: collapse during a free evolution and not during a measurement. So it would be meaningless to ask which theory is better. More details are in Sec. V.}.

The main difficulty in this search is the fact that quantum and classical systems essentially speak different languages and it is hard to make them talk to each other.
The quantum language is based on operators while the classical one---on ordinary functions.
To get around this obstacle, we can try to rewrite quantum mechanics in a more classical form, and thus to obtain a zeroth approximation suitable for our effective theory.

The most convenient  for us is the  approach known as the  complex quantum Hamilton-Jacobi (CQHJ) formulation \cite{lp,john,yang,gdt,poi,chou} which has  attracted growing attention recently.
It is natural to expect that  the new formulation could give us new insights into  fundamental long-standing issues such as the mechanism of quantum collapse.

For brevity and simplicity, we  focus first 
on the simplest yet typical case of the wave function collapse. This case is realised in a one-dimensional Ôsymmetric binary measurementÕ of the momentum where only two outcomes are possible and the 
initial wave function is symmetric with respect to these 
outcomes.  

It can be shown (see Refs.\cite{lp,john,yang,gdt,poi,chou,ign} and Appendix) that
  the  Schr\"odinger equation in this case becomes
\begin{equation}
\label{fin1}
p_t=-\nabla H,
\end{equation}
where
\begin{equation}
\label{def}
p(x,t)=\frac{\hbar}{i}\frac{ \psi'}{\psi},\;\;\; H=V+\frac{p^2}{2m}+\frac{\hat{p}p}{2m}.
\end{equation}
The parity-symmetric wave function before the collapse is the 1:1 superposition of two plane waves with opposite directions:
\begin{equation}
\label{in}
\psi=\frac{1}{\sqrt{2}}(e^{ikx}+e^{-ikx}).
\end{equation}
We need to construct the phenomenological equation that would collapse this wave function into either  $\psi_+=e^{ikx} \;\;\; {\rm or} \;\;\; \psi_-=e^{-ikx}$. 
Following Bohr, we assume that the collapse is caused by the interaction of the quantum system with a semi-classical measuring device. Usually, the interaction is described by the Hamiltonian (or potential) that does not depend on the quantum state. However, such interactions cannot cause the collapse and should be extended to include state-dependent interactions. The phenomenological state-dependent potentials (or their analogues) are familiar in many areas of quantum theory (for instance, equations of Ginzburg-Landau, Gross-Pitaevsky, quantum chemistry etc.). 

A state-dependent interaction is not  a fundamental 
interaction but an approximate one that is obtained, e.g., as a result 
of mean-field approximation which assumes a definite classical value for a mean field. So the proposed approach does not attempt to derive the definite outcomes from a fundamental theory.

The state-dependent potential $V_S$ should be added to the ordinary potential $V$ in Eq. (\ref{fin1})    to obtain:
\begin{equation}
\label{vs}
p_t=-\nabla (H+V_S).
\end{equation}

The gradient of the potential $V_S$ is a more convenient quantity  than the potential itself, and we introduce a special term for it.  It is natural to refer to this gradient as some kind of generalized force; in the context of our problem, we will call it ``the collapsing force'':
\begin{equation}
\label{pot}
F_c=-\nabla V_S.
\end{equation}

The fact that the potential $V_S$ should be state-dependent can be equivalently stated as requirement that the collapsing force $F_c$  should be $p$-dependent.

Due to the shortness of collapse,  the interaction with the apparatus is expected to be dominant during the measurement process, so our Eq. (\ref{vs})  can be written in a simpler form
\begin{equation}
\label{fc}
p_t=F_c.
\end{equation}

To find the form of $F_c$, we invoke the general principles as formulated in the beginning:

(1) Symmetry;

(2) Irreversibility;

(3) Instability.

It is natural to start with the simplest kind of functions, such as low-degree polynomials (which can also be viewed as the first terms of a Taylor expansion):
\begin{equation}
F_c= c_0+c_1p+c_2p^2+c_3p^3+\dots
\end{equation}
According to Eq. (\ref{fc}), the force must be a vector  under parity symmetry ($p\rightarrow-p$) as expected.
Therefore, the  even-power terms in this expansion, $c_0$ and $c_2p^2$,  should be dropped as violating this symmetry.

The requirement that the measurement gives two outcomes,  $p=+q $ or $ p=-q$, translates into the condition that $F_c$ must vanish at $p=\pm q$:
\begin{equation}
F_c(+q)=F_c(-q)=0.
\end{equation}
Therefore, we must have $c_1=-c_3q^2$, and introducing $g=-c_3$ we obtain
\begin{equation}
\label{gp}
F_c=gp (q^2-p^2).
\end{equation}
Consequently, $g$ is interpreted as the effective coupling constant controlling the strength of interaction between the system and apparatus.

In general, $g$ could be a complex number, but in order to satisfy the condition of irreversibility we assume that it is real.

Altogether, our general  equation (\ref{fc}) takes on a `collapsible'   form
\begin{equation}
\label{f}
 p_t=gp (q^2-p^2).
\end{equation}
In addition to Eq. (\ref{pot}), the collapsing force  can be also written in the `potential' form in  `momentum space' with the Higgs-type `collapsing potential':
$F_c=-\partial V_c/\partial p,\;\;V_c=(g/2)(q^2-p^2)^2$ (Fig.1).
\begin{figure}
\includegraphics{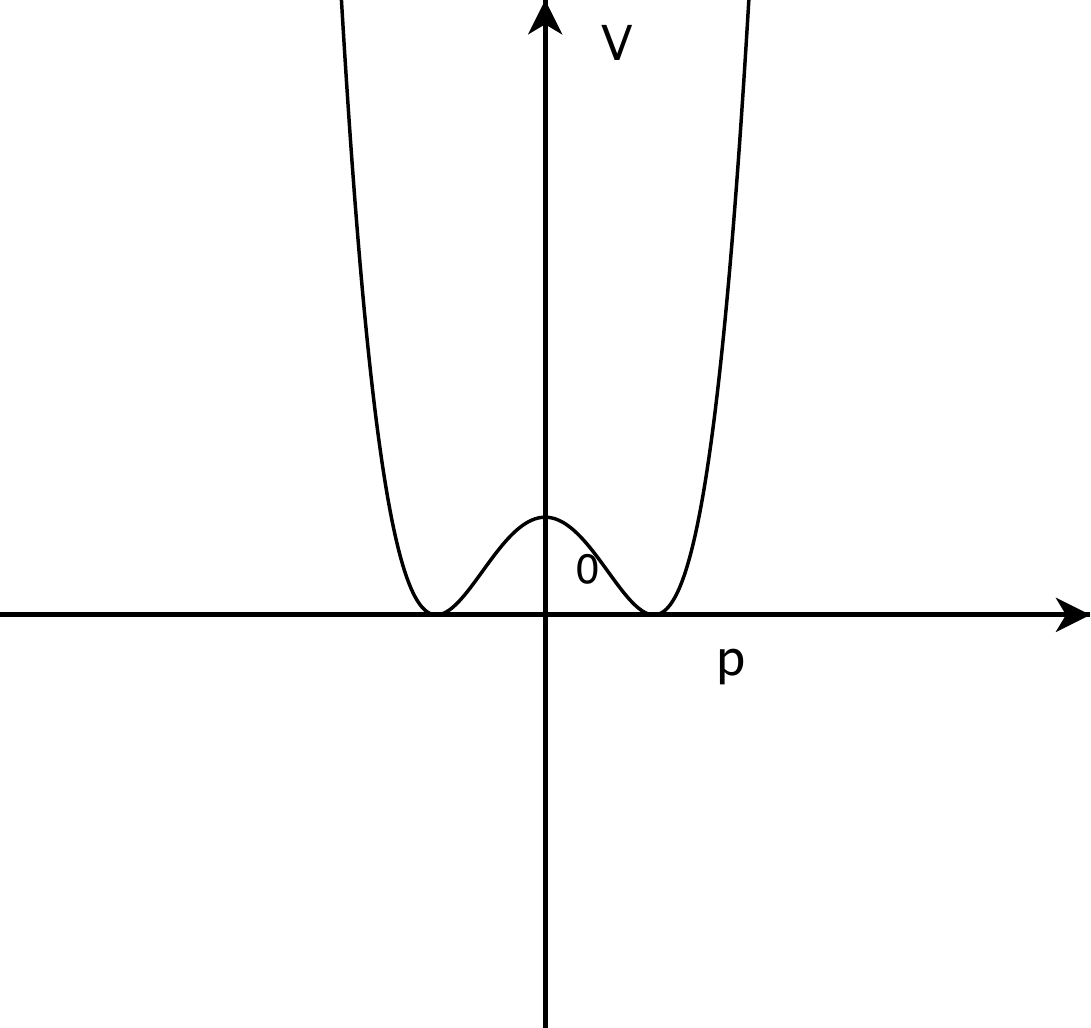}
\caption{{\bf The  schematic form of the `collapsing potential' V(p).} This  2-dimensional plot assumes that $p$ and therefore V are  real numbers. (In general, both are complex so a 4-dimensional plot would be required.) }
\label{sol}
\end{figure}
This illustrates the remarkable  link between our model and the phenomena of  symmetry breaking, phase transitions, and synergetics. The collapsing potential can be compared to the `mean field' and the measured quantity---to the `order parameter'.

Solving Eq. (\ref{f}) with the initial condition (\ref{in}),  
yields 
\begin{eqnarray}
\label{a1}
p(x,t)=  \frac{qp_0(x)B(t)}{\sqrt{p_0^2(x)[B^2(t)-1]+q^2}},\\\label{a2}
\;\;
B(t)=\exp{(gq^2t)}, \;\;\;  p_0(x)=iq \tan{kx}.
\end{eqnarray}

Finally, using the definition (\ref{def})   we can determine the {\em continuous} behavior of the wave function {\em all the way through the measuring process}---in other words, {\em the process of the wave function  collapse}:
\begin{equation}
\label{a3}
\psi(x,t)= \frac{B(t)\cos{kx}+\sqrt{1-B^2(t)\sin^2{kx}}}{B(t)+1}N(t),
\end{equation}
where $N(t)$ is the normalization factor.
In particular, we obtain the following formula for the characteristic timescale $\tau_c$ of the  collapse:
$\tau_c\sim 1/(gq^2)$.
 
 Asymptotically, when $B(t) \gg 1$, the wave function (\ref{a3}) can take one of the two forms (in each domain of length $\pi/2k$):
 \begin{equation}
 \label{a3a}
\psi_+=e^{ikx} \;\;\; {\rm or} \;\;\; \psi_-=e^{-ikx}.
\end{equation}
This two-valuedness arises due to the two-valuedness of the square roots in Eqs. (\ref{a1},\ref{a3}). 
 An interesting question now is: How does the wave function `know'  which way to collapse: $\psi_+$ or $\psi_-$?  
The answer is that  if the initial wave function has {\em the exact} form (\ref{in}), then {\em it does not know,} i.e., the two-valued ambiguity in (\ref{a3a}) {\em cannot be resolved.} 
 
This is because the initial $p_0$ is {\em purely imaginary.} If, on the other hand, $p_0$ has a non-zero real part $Re\; p_0$, then the definite outcome arises: In case of positive real part, $Re\; p_0>0$, the asymptotic wave function is $\psi_+=e^{ikx}$; if the real part is negative, $Re\; p_0<0$, the asymptotic wave function is $\psi_-=e^{-ikx}$.
 Thus the measurement outcome is completely determined by the  sign of the real part of $p_0$.
It is natural to presume that $Re\;p_0$ can take up positive and negative values with equal probability as required by the Born rule.

We emphasize that only {\em infinitesimal} non-zero values of $Re\; p_0$ are required for the emergence of a definite outcome in the scheme based on the simple equation (\ref{f}). 

A question can be raised whether $Re\; p_0$ can be considered a hidden variable, the dependence on which of the outcome is a step function.

Using a step function to represent the outcome dependence on $Re\;p_0$ leads to the following problems:
If the dependence of the outcome on $Re\;p_0$   is represented by a step function, the domain  of this function should be infinitesimally small. If finite values of  $Re\;p_0$ are allowed to be included in that domain, then these values would be incompatible with the state described by the wave function (\ref{in}).  But if we consider other states compatible with finite values of  $Re\;p_0$, then such a step function would not be a correct predictor of the outcome.
Therefore, the outcome dependence on $Re\;p_0$ has a more singular character than the step function. By the same argument, the outcome dependence on 
$Re\;p_0$ cannot be represented by {\em any} function with finite domain. Also, it cannot be represented by a discrete binary variable $\lambda$ taking on values $\pm 1$ because in that case the dependence of this `stand alone' variable on $Re\;p_0$ would be lost.

For these reasons we prefer not to call $Re\;p_0$ a hidden variable.

The real issue, though, is not so much in the name but in a  possible contradiction with various no-go theorems which have been formulated in relation to theories with hidden variables of various sorts. These issues are discussed in Sec. V.F where it is shown that no contradictions arise.

\section{Interpretation} 
A few remarks are in order concerning  the interpretation of our `collapsible equation', Eq.  (\ref{f}).

(a) It is not supposed to be  a fundamental equation describing exactly the collapse process in every detail. Rather, it is a phenomenological, effective approximation which is meant to capture the main features of the collapse: instability, irreversibility, and symmetry breaking, which could lead to observable consequences such as the finite time-scale of the collapse process.
 
In any case, the construction of more elaborate models of the wave function collapse seems to require more experimental data than is available at the moment.

(b) In contrast with the Bohm-Bub models of explicit collapse \cite{bb}, Eq. (\ref{f}) does {\em not} contain hidden variables.

There is a large number of works that {\em replace} the Schr\"odinger equation with a non-linear one (see, e.g., \cite{bbm, wn}). In contrast, our collapsible equations {\em complement} the Schr\"odinger equation. They have  a different mathematical structure and distinct observational consequences.

(c) The ``superluminality'' argument which is often invoked to limit the use of non-linear equations in quantum mechanics does {\em not} apply in our case for the reasons discussed in Sec. V.  

\section{Experimental consequences} 
As with any phenomenological equation, care is required in comparing its predictions with experiment.  Although our derivation allows one to establish some necessary conditions of its validity, these conditions may not be sufficient, and only experiment can decide on this issue.
First of all, a certain balance between the characteristic times involved should be observed. Ideally, one would aim at the interaction-switching time much less than the collapse time, and the collapse time much less than the `energy' time-scale $\hbar/\tau_E$ where $E$ is the characteristic energy scale involved in a particular experiment.

One scheme of an experiment could look as follows. The initial state is created, then the measurement interaction is switched on at the instant $t=0$, then the measurement result is read-off. In the conventional picture, the interval between the   second and third events can be as small as possible (it is bounded from below by the resolution time of the detector). However, in our picture, the collapse time cannot be less than $\tau_c$. Therefore, if the time-scale $\tau_c$ exceeds the resolution time-scale, the theory can be tested.

To compare the sensitivities of different kinds of experiments, it is convenient to introduce the dimensionless collapse time $\kappa=\tau_c/\tau_E$. 
Except for the pioneering work of Papaliolios \cite{pap} back in 1967, no dedicated searches for the finite time of the collapse have been performed (to my knowledge). So we attempt to obtain rough estimates on the allowed values of $\tau_c$ and $\kappa$ from various other experimental results assuming that null effects were observed in them on a time scale of $\tau_m$ which is the shortest time scale probed in a specific experiment (Table I).

\begin{table*}[htdp]
\caption{\bf Experimental constraints on the collapse time $\tau_c$.}
  \begin{ruledtabular}
\begin{tabular}{ccccc}
Experiment&Ref.&$\tau_m$ (s)& $\tau_E$ (s)& $r$\\ \hline
Photon polarization&\cite{pap}&$ 7.5\times 10^{-14}$&$\sim 10^{-15}$&$\sim 10^2$\\
Neutron interferometry&\cite{gkz}&$2.7\times 10^{-2}$&$2.3\times 10^{-12}$&$10^{10}$\\
Quantum jumps&\cite{nsd}&$1$&$\sim10^{-15}$&$\sim10^{15}$\\
Nonlinearity test&\cite{bhi}&$1$&$\sim10^{-9}$&$\sim10^{9}$\\
Femtosecond optics&\cite{blr}&$\sim10^{-13}$&$\sim 7\times  10^{-17}$&$\sim10^{3}$\\
Bose-Einstein condensate&\cite{dcc}&$\sim10^{-4}$&$1.8\times  10^{-3}$&$\agt 5\times  10^{-2}$\\
EPR correlations&\cite{zbg}&$5\times  10^{-12}$&$\sim10^{-15}$&$5\times  10^{3}$\\
\end{tabular}
\end{ruledtabular}
\label{default}
Absolute upper bound: $\tau_c\alt\tau_m$. Relative upper bound: $\tau_c/\tau_E\alt  r$. All figures are to be taken as order-of-magnitude estimates rather than rigorous limits.
\end{table*}

It is natural to start with the experiments \cite{gkz,bhi} that tested specific non-linear versions of the Schr\"odinger equation \cite{bbm,wn, mdo}. However, they turn out {\em not} to be the most sensitive for our purposes.
The well-known observation of `quantum jumps' \cite{nsd} is another natural place to look for constraints, but again they are found to be rather weak. 
Next, we turn to the ultrafast measurements using femtosecond  pulses of laser light \cite{blr}. 

Here, the limits begin to improve.  
Finally, an excellent source of constraints appears to be experiments with Bose-Einstein condensates.
The 
observed dynamics
turned out surprising in many respects, and it would be  interesting to see if the idea of a finite-time collapse could be helpful there.

This area is virtually unexplored, and there is not one, but many promising directions of attack.  Consequently, without attempting to be exhaustive, we restrict ourselves to outlining just two general strategies. 

One approach is to look for the longest characteristic times $\tau_E$, comparable or better than those achieved in the BEC experiments. For example, the distances between the energy levels of ultra-cold  neutrons bouncing from a horizontal mirror in the earth gravity field are of the order of $ 10^{-12}$ eV  (or peV) \cite{dur} which corresponds to $\tau_E \sim 10^{-3} $ s. Because times, such as the time of flight, can be measured with much better accuracy, this type of experiment (suggested for other purposes) could be a useful starting point for a test proposal.

Alternative strategy would be to start from the experiments with the shortest possible times $\tau_m$, such as those involving ultrashort  pulses. This is a rapidly evolving area progressing from the  femtosecond scale  to atto-, zepto-, and even yoctoseconds ($10^{-24}$ s).

We emphasize that if the theory is confirmed that would not mean at all that the conventional quantum mechanics is refuted. It would only mean that the approximation of the instantaneous collapse is too crude and should be replaced by a more realistic approximation of a finite time-collapse.  As was mentioned in the introduction, the assumption of  instantaneous collapse is just an approximation rather than a fundamental principle of quantum theory, and the validity of this approximation does not affect the validity of quantum theory as a whole.

Similarly, if our specific Eq. (\ref{f}) is experimentally excluded, that would not mean that the whole approach is invalidated, because a variety of different equations could be built on the same conceptual basis which could lead to different experimental predictions.

\section{Discussion}
In constructing the collapse-enabled  Schr\"odinger equations, our method is based on three well-established principles:

1. {\em Symmetry breaking.} The binary symmetric measurements on one particle in the absence of decoherence are the fundamental prototype measurements that should be studied first, both theoretically and experimentally.

Furthermore, the collapsible equations for such measurements should be constructed phenomenologically using the minimal set of most general assumptions. 
Specifically, we single out the following two conditions that are widely accepted as reliable, non-controversial characteristics of the collapse process:

2. {\em Irreversibility.} This is  generally acknowledged as the most important, essential feature of the collapse. 

3. {\em Instability} (or amplification). The  key role of these principles in the theory of measurement was succinctly formulated by Bohr as follows:``...every atomic phenomenon is closed in the sense that its observation is based on registrations obtained by means of suitable {\em amplification} devices with {\em irreversible} functioning such as, for example, permanent marks on the photographic plate caused by the penetration of  electrons into the emulsion.'' (italics added---A.I.). In the earlier literature, the same principles were often described in the language of  ``uncontrollable perturbations'' arising when the quantum system interacts with the measuring device.

We note that the requirement of finite time is {\em not} among our basic principles. Rather, it comes out as a natural consequence of them.

Our method reflects the fact that decades of concerted efforts towards clarifying the fundamental mechanisms of collapse and the measurement process have met with limited success. This means that the collapse problem is an extremely challenging as well as an highly interdisciplinary one. It involves several distinct areas of theoretical physics and modern mathematics.  As a consequence, there is little hope of progress unless one uses the phenomenological approach first and ask questions about first principles later.

Wave function collapse can be regarded as a `macroscopic quantum phenomenon'. Therefore we can  turn for inspiration to the successful theories of other macroscopic quantum phenomena such as superfluidity and superconductivity.

In particular, it would be interesting  here to draw a parallel with the puzzle of superconductivity, which took almost 50 years to unravel and it is still at the forefront of physics research. 
This ultimate success was based on the consistent use of the phenomenological, effective-theory type of approaches before embarking on the final quest for a microscopic theory. The same strategy could be worth trying in the collapse problem.

The analogy can be continued in another respect. The phenomenological theory of superconductivity of Ginzburg and Landau uses a non-linear equation for the  wave function. This does not mean, however, that the superposition principle is violated. Likewise, our use of a non-linear collapsing equation does not necessarily mean the violation of that principle because it is meant to be understood as a description at the phenomenological level which is  not necessarily fundamental. 

Many alternative models can be constructed  by choosing different representations,   forms of the collapsing force and/or adding hidden variables to the fundamental general equations belonging to the family exemplified by Eqs. (\ref{vs}) and (\ref{f}).
However, given the probable complexity of the collapse phenomenon, it would be desirable to focus  first (both theoretically and experimentally) on the  most fundamental, the ``purest'' form of collapse---the binary symmetric measurement.

With suitable modifications, our methods could be applied in other subfields of quantum theory where a non-potential generalization of the Schr\"odinger equation is required by the conditions of the problem. Examples include the theory of dissipative systems, motion under the forces of radiative friction, etc. More generally, one can expect that interesting  connections with nanoscience, quantum gravity and quantum cosmology could also be possible. 
 
\subsection{Schematic of main finding}

Quantum state collapse is represented as a kinetic process of irreversible symmetry breaking driven by the instability of the initial state (Fig. 1-3).
\begin{figure}
\includegraphics{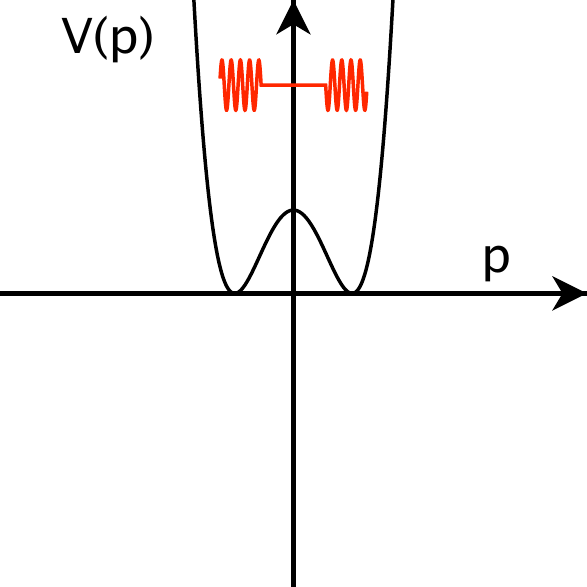}
\caption{The initial wave function is a symmetric superposition of opposite momentum values $p=+q$  and  $p=-q$.}
\label{S1}
\end{figure}
\begin{figure}
\includegraphics{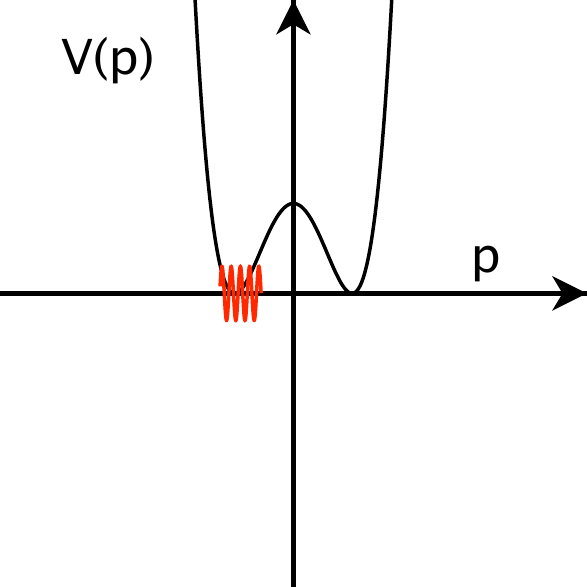}
\caption{After the measurement interaction  gave the result $p=-q$, the wave function collapses to $\psi_{-}=exp(-iqx)$.}
\label{S2}
\end{figure}
\begin{figure}
\includegraphics{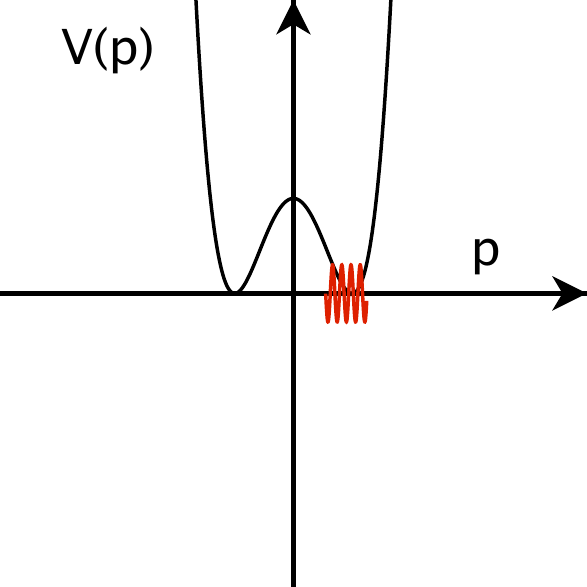}
\caption{After the  measurement interaction  gave the result $p=+q$, the wave function collapses to $\psi_{+}=exp(+iqx)$.}
\label{S3}
\end{figure}

\subsection{Physical ideas underlying our framework}
Our approach and results may look deceptively simple and questions could 
 be raised as to what makes this possible.

The explanation is  that all the hard work is done by combining six powerful insights 

from several fronts 

of  theoretical physics, which have not been put together before:

-complex quantum Hamilton-Jacobi formulation of quantum mechanics

-principle of  symmetry breaking 

-the effective theory approach

-the idea of irreversibility

-the measured quantity as the order parameter

-Bohr's insight that quantum phenomena occur due to irreversible amplifying interaction of a classical and quantum systems.
\subsection{Where do we stand in terms of various interpretations?}
Although our formulation is designed to be interpretation-neutral, it can also be seen as the logical development of the conventional Copenhagen interpretation.
As discussed in the main text, those who favour the statistical ensemble interpretation of quantum mechanics would probably be skeptical about our approach. In the spirit of ``experimental metaphysics'', we believe that experimental tests may help to clarify or update one's philosophical position.

The same can be said about the Many-Worlds interpretation. In fact, experimental searches for collapse dynamics are needed from the standpoint of {\it any} interpretation because the results of such experiments will help to decide which one is closer to Nature. 

If these searches fail to produce evidence for collapse, this would be a serious argument in favour of the collapse-denying interpretations, including Many-Worlds and statistical. Such arguments are hard to achieve from just theoretical reasoning. 

Also,  it is well-known that in some situations the wave function is `more real' than in others (superconductivity, Bose-Einstein condensation etc.). Consequently, collapse may be a more real phenomenon in those situations.
In any case, the wave function satisfying our collapsible equation should be interpreted as an `effective' or `interpolating' wave function rather than a `genuine' wave function which ceases to exist once the system starts to interact with the measuring device, according to the standard (von Neumann's) model of the measurement process.

\subsection{Comparison with the approaches of Bohr and von Neumann}
In any treatment of the measurement problem
the question arises: is the measuring apparatus to be treated as a quantum system or not? On this crucial point, the opinions of quantum theorists appear to be split.
One school of thought (Bohr, Heisenberg, Dirac, and Born) accepts Bohr's idea that   the measuring device should be described approximately as classical and not quantum-mechanical. Consequently, the wave function of the apparatus is not introduced and the Schr\"odinger equation for the combined set ``system plus apparatus'' is not discussed or solved. 
The alternative school (von Neumann, Landau and Peierls, Pauli, and most modern textbooks) assumes that the measuring device should be treated on the same footing as any other quantum system. As a consequence, most modern treatments introduce the wave function of the apparatus as a matter of fact that does not require further justification.

At the moment, one of the fruitful if challenging issues in quantum mechanics is the question whether  the superposition principle is universally valid or not. Obviously, this is just another way to ask the same question.
Again, this problem is attacked not by purely philosophical discussion, but in the spirit of `experimental metaphysics'.  However, despite impressive progress in the area, it would be fair to say that the final solution is not yet on the horizon.

We do not introduce explicitly the wave function of the measuring device. This is in accord with Bohr's ideas that  the the measuring device is best
described approximately and classically rather than quantum-mechanically. In fact, Bohr appears to 
have reservations regarding  the crucial step of ``quantizing the apparatus'' (first made by  von Neumann) as he did not  discussed it in his published works.
From the modern perspective, this may be viewed 
as a flaw or even a serious defect of such an interpretation.
On the other hand, Bohr's reservation
might  probably  be justified because to this day,  the efforts spent in quantizing the apparatus have not been completely successful. Moreover, this problem is considered fundamental but intractable in the forseeable future. 
For example, 
it was pointed out \cite{bh} that no observer can write down his own wave function just because the human brain capacity is not large enough for that. A similar type of argument goes for an inanimate device as well.

We, of course, do not pretend to solve this part of the problem. Instead, we go around it and concentrate on the behaviour of the quantum system rather than the measuring device. In this way, we act in the spirit of  `effective' theory, which is well-known and extremely fruitful in the area of quantum fields  (see, e.g., \cite{wei}), but is less appreciated in the context of quantum mechanics. 
The essence of the effective theory approach is to integrate out all unknown degrees of freedom and focus on what is essential and observable.

\subsection{Relation to decoherence}
Although the theory of decoherence made an important step towards solving  the collapse problem, it has not solved it.

Our paper does not include decoherence explicitly, but this is not because it is irrelevant. The real reason is that the collapse puzzle is a difficult and unsolved task. The common approach of a theoretical physicist in such cases is to tackle first  an idealized setting where decoherence is ignored. 

Moreover, this simple setting could be much more realistic than it seems at first sight. Indeed, we could imagine physical conditions under which the effects of decoherence are greatly reduced, such as very low noise, low temperature and similar  types of environment.  In this `very clean' environment we expect that decoherence will be greatly affected, but the collapse will not. Consequently, analysis of collapse in the absence of decoherence is not merely an academic exercise, but a viable physical model for  specific environmental conditions.

\subsection{The Einstein-Podolsky-Rosen set-up and no-go theorems}

In this paper, we do not treat any EPR-related issues leaving them for further work. We feel that the phenomenon of collapse, if real, could be complicated and not necessarily universal for all types of systems and experiments.  One may even wonder if the quantum collapse could be as complex as its gravitational cousin. In any case, it seems prudent to approach this difficult problem in a step-by-step manner starting with the simplest possible set-up: a measurement made on one non-relativistic particle.

As Braginsky and Khalili note in Ref. \cite{bh}  (p. 28): ``The presence of two or more degrees of freedom can change the character of the measurement substantially''.

Perhaps, to emphasise the differences, it could be helpful  to introduce special terms like, for example, ``contact''  and ``non-contact'' collapse, the latter describing the EPR-type situations.

We also note that under a different set of assumptions, an EPR-type experimental search for a finite-time collapse was recently proposed  in Ref. \cite{MP}.

Because of the fact that our simple model (\ref{f}) treats only  the case of one particle and two measurement outcomes, it avoids tensions with Bell's inequalities \cite{bell}  as well as various other no-go theorems due to  Gleason \cite{glea}; Kochen and Specker \cite{ks};  Conway and Kochen  (or ``the Free Will theorem'') \cite{ck}; Colbeck and Renner \cite{cr}; Pusey, Barrett and Rudolph \cite{pbr}.

More specifically,  both Gleason's and  Kochen-Specker's theorems require that the dimensionality of Hilbert space be greater than two \cite{glea,ks}---which does not apply in our case (\ref{f}). 

The Free Will theorem, in both original and strengthened formulations, requires a two-particle state due to its TWIN axiom \cite{ck}. Therefore, it does not clash with our model (\ref{f}) either.

Colbeck and Renner  \cite{ck} extended Bell's prohibition of local hidden variables  by considering  hidden variables that have both local and global components and ruling them out as well. Again, our model (\ref{f}) is immune from this exclusion theorem because it (the theorem) requires at least a two-particle system, among other things \cite{cr}.



Pusey, Barrett and Rudolph \cite{pbr} have recently proved another no-go theorem excluding a certain class of theories with hidden parameters $\lambda$.
The following quote from Ref.  \cite{pbr} summarizes their work nicely: ``Our main result is that for distinct quantum states 
$|\psi_0\rangle$ and $|\psi_1\rangle$, if the distributions $\mu_0(\l)$ and $\mu_1(\l)$ overlap (more precisely: if $\D$, the intersection of their supports, has non-zero measure) then there is a contradiction with the predictions of quantum theory.''

It follows immediately that this result does not apply to our model (\ref{f}) because for distinct quantum states the distributions of our infinitesimal  parameters do not overlap.


\subsection{Non-linear quantum mechanics and superluminality}
It has been proposed by Gisin \cite{gis} among others that non-linear modifications of quantum mechanics could lead to faster than light communication. However, several points have to be taken into account here:

1) Our collapsible Schr\"odinger equations are {\em not} of the type considered in Ref. \cite{gis}. Our non-linearity arises {\em only} at the measurement stage, while equations studied in Ref.  \cite{gis} are non-linear {\em always}. For this reason, superluminal communication is not made possible by our equations and the objection does not apply.

2) For completeness, we also mention that in subsequent papers \cite{fss,kent,bgh} it was pointed out that the argument of Ref.\cite{gis} is not free from possible loopholes. However, due to the above point, we do not need to appeal to these loopholes in order to demonstrate that our equations are free from the superluminality objection.


\subsection{Is consciousness needed for collapse?} 
In our approach, an observer or the consciousness is not required for collapse to occur (cf. \cite{tha}). Moreover, our experimental upper limits on the collapse time (Table 1) show that this process is too fast for human consciousness which typically operates on much longer time-scales.

\subsection{Collapse and relativity}
Relativistic  generalisations of the collapse concept are notoriously difficult and are not discussed here. We would like to stress, however, that at the moment  the proposed collapse mechanism does not appear to be incompatible with relativity. On the contrary, a relativistic generalisation of our collapsible equations seems a worthwhile direction to pursue. 

\subsection{de Broglie-Bohm formulation}
Some readers may get an impression that we are using or advocating some version of the  de Broglie-Bohm (dBB) interpretation of quantum mechanics \cite{hol}. 
We emphasise that this is {\em not } the case \footnote{This phrase should not be construed as an implicit criticism of such interpretations.}.  What we do is just borrowing a useful form of rewriting the Schr\"odinger equation. This can be considered as a purely mathematical transformation that is completely neutral with regard to interpretational issues. The dBB interpretation involves much more than such a rewriting, e.g., particle trajectories. These extra features are not necessary in our approach.

\subsection{Other types of measurements}
Today we know many more  types of measurements than  in 1930s. 
Examples include weak, indirect, non-demolition, continuous, interrupted and other varieties of measurements. We emphasise that all these are beyond the scope of the present paper. 

Our focus here is only on the simplest, ``canonical'' kind of measurement, which was also historically one of the first examples where the measurement problem was born.  Without a thorough understanding of this example, there is little chance of progress in the more intricate cases. 

Moreover, it would be perhaps over-optimistic to expect that one and the same theory can describe {\em all} kinds of measurements. (It would be similar to expecting, for instance, that one and the same theory can describe both  strong  and gravitational interactions.)

\subsection{Theories of dynamical collapse}
On a first view, our framework can be seen as a competitor to the well-known theories that represent collapse as a continuous dynamical process \cite{bas}. In fact, it is not. The main reason is that the two approaches set different goals from the start even if the term ``collapse'' happens to figure prominently in both of them.

Briefly, this term is used in  two rather different senses so one  wonders if two different terms would be more appropriate here.
Our collapse  has been induced by the measurement  (``M-collapse'').
 In  the  theories of dynamical collapse, it is occurring spontaneously, during the free evolution (``S-collapse'').  This is a considerable difference with far-reaching consequences.
 
 The  theories of dynamical collapse do not address the mechanism of M-collapse. Likewise, we do not discuss the S-collapse. We ask the question: what happens to the wave function during the measurement? The  theories of dynamical collapse ask:   Why do the wave functions not spread out so much that the classical picture becomes blurred (problem of macro-objectification)?
 
 It might be possible that in the end, the answers to both questions could come from one and the same ultimate source. However, for all similarities between the two problems  there seem to be as many differences, so it would be reasonable to treat them as separate issues for the time being.

Our model and the dynamical collapse theories are made for two different purposes; they have different mathematical structure and different observational consequences.

\subsection{Meaning of collapsing force}
Introduction of the collapsing force (Fig. 4) is the main  phenomenological element of our program.
\begin{figure}
\includegraphics{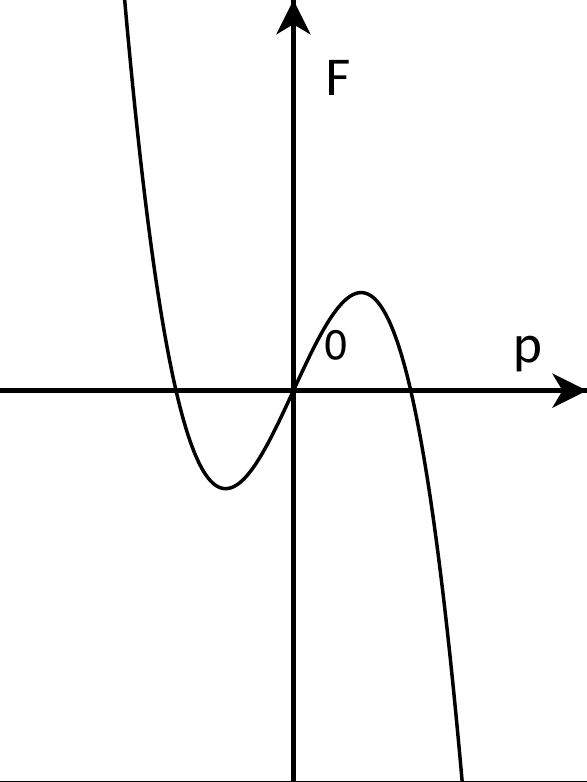}
\caption{{\bf The schematic form of the `collapsing force' F(p).} This  2-dimensional plot assumes that $p$ and therefore F are  real numbers. (In general, both are complex so a 4-dimensional plot would be required.)}
\label{sol}
\end{figure}
 As usual in such cases, we do not attempt to derive it from first principles, but rather postulate it on the combined strength of a series of  independent arguments:

---The irreversible collapse during a measurement can be viewed as sort of relaxation process. Then the appearance of a collapsing force is naturally understood as the emergence of a relaxation-driving dissipative force without which relaxation would be impossible.

---In the same vein, the kinetic mechanism of collapse invites a parallel with the classical kinetic theory. The collapsing force, then, plays a similar role to the collision term in the Boltzmann equation.



---The task of constructing the collapsible Schr\"odinger equation can be formulated as the problem of finding a {\em smooth interpolation} between the uncollapsed and collapsed wave functions. A new term that is needed for such an interpolation would be equivalent to  a collapsing force.

---What arguments force us to assume that the collapse occurs instantaneously, given that we are not aware of any other instantaneous processes in Nature? 

---If, on the other hand, the collapse is not instantaneous but takes a finite time, then it seems  plausible for purely mathematical reasons that one or another kind of extra terms (`generalised collapsing forces') should be introduced into the CQHJ-Schr\"odinger equation to produce this finite time.

---Within the CQHJ approach, the Schr\"odinger equation acquires a form so similar to  Newton's second law that it almost `begs for'  introducing 
such a new force and  exploring its consequences.

\subsection{Is quantum mechanics to be modified?}
Do our equations require a modification of standard quantum mechanics?  At first sight, the answer to this would be ``Yes'', but we believe that such a conclusion would be premature and unwarranted.

As was mentioned in the main text, our goal was not to modify quantum mechanics, but to obtain effective equations for the collapse process.
The fact that our equations are non-linear, and the Schr\"{o}dinger equation is linear is not enough, in our view, to claim that  these equations violate quantum mechanics.


The reason is that the Schr\"{o}dinger equation is exact, and our equation is approximate. But it is well-known that the exact and linear equation (Schr\"{o}dinger) can lead to approximate non-linear ones when suitable approximations are made (equations of Hartree-Fock, Gross-Pitaevsky etc.).

Indeed, as emphasized by Fock \cite{foc}, in quantum theory the distinction between ``fundamental'' and ``effective'' concepts becomes a matter of convention because approximations play the principal, not a secondary role. For example, from the perspective of quantum field theory, the Schr\"{o}dinger equation itself becomes an approximate ``effective'' equation, rather than  a ``fundamental'' law.
Finally, from  purely experimental perspective, the `fundamental or effective' question is hardly the most 
pressing one.
\section{Conclusions and outlook}
To sum up, a new class of quantum-mechanical non-linear equations for the wave function collapse has been proposed
 that  combines conceptual simplicity  and predictive power.
 Also, we have  reported a set of constraints and outlined future opportunities for the sensitive experimental tests.

It is exciting to realize that only two or three orders of magnitude separate us from deeper, dedicated  probes of the wave function collapse. The results will help to clarify both foundational and practical issues in quantum mechanics such as the status of the Many-Worlds interpretation and the ultimate limits on the speed of quantum computation and information processing.

{\em Acknowledgments.} The author  thanks  M. Albrow, R. D\"orner, L.  Ignatieva,   V. A. Kuzmin, J.-P. Magnot, M. E. Shaposhnikov, and F. Thaheld  for taking time for  reading, valuable comments and discussions.
\appendix*
\section{Complex quantum Hamilton-Jacobi equation}
To make the article self-contained, we provide below a 
derivation of the complex quantum Hamilton-Jacobi equation.

Let us start with the ordinary Schr\"odinger equation for a particle of mass $m$ in the potential $V({\bf r})$:
\begin{equation}
\label{s}
i\hbar\psi_t=-\frac{\hbar^2}{2m}\Delta \psi +V\psi,
\end{equation}
and
 introduce a new variable ${\bf p}({\bf r}, t)$:
\begin{equation}
\label{p}
{\bf p}=\frac{\hbar}{i}\frac{\nabla \psi}{\psi}.
\end{equation}
Our goal   is to obtain a closed dynamical equation for ${\bf p}$.
Differentiating Eq. (\ref{p}) with respect to time, we find:
\begin{equation}
\label{4s}
{\bf p}_t=\frac{\hbar}{i}\frac{\nabla \psi_t}{\psi}-{\bf p}\frac{ \psi_t}{\psi}.
\end{equation}
It is convenient to start $\psi$-elimination from the second term of Eq. (\ref{4s}). For this purpose, we first  rewrite $\Delta \psi$ as
\begin{equation}
\label{6s}
\Delta  \psi=\nabla\left(\frac{i}{\hbar}{\bf p}\psi \right)=\frac{i}{\hbar}\left(\psi\nabla\cdot{\bf p}+{\bf p}\nabla\psi\right).
\end{equation}
Inserting this into the Schr\"odinger equation (\ref{s}) and multiplying both sides by $i{\bf p}/(\hbar \psi)$, we obtain:
\begin{equation}
\label{11s}
-{\bf p}\frac{ \psi_t}{\psi}=\frac{1}{2m}{\bf p}\nabla\cdot {\bf p}+\frac{i}{2m\hbar}{\bf p} p^2+\frac{i}{\hbar} {\bf p}V.
\end{equation} 
To eliminate $\psi$ from the first term of Eq. (\ref{4s}), we substitute the right-hand side of Eq.  (\ref{6s}) into the Schr\"odinger equation (\ref{s}) instead of $\Delta \psi$, then take the gradient of both parts and divide both of them by $\psi$, which yields:
\begin{eqnarray}
\label{13s}
\nonumber -i\hbar \frac{\nabla \psi_t}{\psi}=\frac{i\hbar}{2m}\nabla(\nabla\cdot {\bf p})-\frac{1}{2m}{\bf p}\nabla\cdot {\bf p}-\\\frac{1}{2m}\nabla  p^2-\frac{i}{2\hbar m}p^2 {\bf p}-\nabla V-\frac{i}{\hbar}{\bf p} V.
\end{eqnarray}

We now add  this formula, part-by-part, to Eq. (\ref{11s}) and then use  Eq. (\ref{4s}) to finally obtain:
\begin{equation}
\label{fin}
{\bf p}_t=-\nabla V-\frac{1}{2m}\nabla  p^2+\frac{i\hbar}{2m}\nabla(\nabla\cdot {\bf p}).
\end{equation}

So the complete elimination of the wave function is indeed possible, and Eq. (\ref{fin}) reproduces the dynamical equation  that 
 plays the role of the Schr\"odinger equation in the CQHJ approach.  
 A slight rearrangement gives it  an elegant shape:
\begin{eqnarray}
\label{fin1}
{\bf p}_t=-\nabla H,\\
{\hat{\bf p}}=-i\hbar \nabla, \;\;\;
H=V+\frac{p^2}{2m}+\frac{{\hat{\bf p}}{\bf p}}{2m}.
\end{eqnarray}


\clearpage

\end{document}